\documentclass[a4]{article}

\usepackage[frenchb,english]{babel}
\usepackage{fancyvrb}
\usepackage[applemac]{inputenc}
\usepackage{a4}
\usepackage{graphics}
\usepackage{amsthm}
\usepackage{amstext}
\usepackage{amsmath}
\usepackage{amsbsy}
\usepackage{multirow}
\usepackage{amssymb}
\usepackage{amsfonts}
\usepackage{wasysym}
\usepackage[vflt]{floatflt}
\usepackage{epsfig}
\usepackage{subfigure}
\usepackage{rotating}
\usepackage{hyperref}
\newcommand{\ANKARA}      {3}
\newcommand{\ANNECY}      {14}
\newcommand{\BARI}        {35}
\newcommand{\BARIINFN}    {21}
\newcommand{\BERN}        {6}
\newcommand{\BOLOGNA}     {15}
\newcommand{\BOLOGNAINFN} {16}
\newcommand{\BRUSSELS}    {38}

\newcommand{\DUBNA}      {18}
\newcommand{\FRASCATI}   {17}
\newcommand{\FUNABASHI}  {30}	
\newcommand{\HAIFA}      {28}
\newcommand{\HAMBURG}    {24}
\newcommand{\GAZWADONG}  {32}
\newcommand{\KARIYA}     {33}
\newcommand{\KOBE}       {5}
\newcommand{\LAQUILA}    {22}
\newcommand{\LNGS}       {19}
\newcommand{\LYON}       {7}
\newcommand{\MOSCOWINR}  {1}
\newcommand{\MOSCOWITEP} {25}
\newcommand{\MOSCOWLPI}  {9}
\newcommand{\MOSCOWSINP} {4}
\newcommand{\MUNSTER}    {26}
\newcommand{\NAGOYA}     {27}
\newcommand{\NAPOLI}     {20}
\newcommand{\NAPOLIINFN} {2}
\newcommand{\PADOVA}     {13}
\newcommand{\PADOVAINFN} {11}
\newcommand{\ROMA}       {34}
\newcommand{\ROSTOCK}    {37}
\newcommand{\SALERNO}    {12}
\newcommand{\STRASBOURG} {23}
\newcommand{\TUNIS}      {10}
\newcommand{\URBINO}     {29}
\newcommand{\UTSUNOMIYA} {36}
\newcommand{\ZAGREB}     {31}
\newcommand{\ZURICH}     {8}
\newcommand{\CORR}       {*}
\newcommand{\NOWASAN}     {f}
\newcommand{\NOWBERN}     {h}
\newcommand{\NOWFRASCATI} {e}
\newcommand{\NOWINAFIASF} {b}
\newcommand{\NOWLNGS}     {a}
\newcommand{\NOWPUSAN}    {d}
\newcommand{\NOWROMA}     {c}
\newcommand{\NOWSUBATECH} {g}

\newcommand{\OperaInstitutes}{
\MOSCOWINR   . INR-Institute for Nuclear Research of the Russian Academy of Sciences, RUS-117312 Moscow, Russia \\
\NAPOLIINFN  . INFN Sezione di Napoli, I-80125 Napoli, Italy \\
\ANKARA      . METU-Middle East Technical University, TR-06531 Ankara, Turkey \\
\MOSCOWSINP  . SINP MSU-Skobeltsyn Institute of Nuclear Physics of Moscow State University, RUS-119992 Moscow, Russia \\
\KOBE        . Kobe University, J-657-8501 Kobe, Japan \\
\BERN        . Albert Einstein Center for Fundamental Physics, Laboratory for High Energy Physics (LHEP),
               University of Bern, CH-3012 Bern, Switzerland \\
\LYON        . IPNL, Universit\'e Claude Bernard Lyon 1, CNRS/IN2P3, F-69622 Villeurbanne, France \\
\ZURICH      . ETH Zurich, Institute for Particle Physics, CH-8093 Zurich, Switzerland \\
\MOSCOWLPI   . LPI-Lebedev Physical Institute of the Russian Academy of Sciences, RUS-117924 Moscow, Russia \\
\TUNIS       . Unit\'e de Physique Nucl\'eaire et des Hautes Energies (UPNHE), Tunis, Tunisia \\
\PADOVAINFN  . INFN Sezione di Padova, I-35131 Padova, Italy \\
\SALERNO     . Dipartimento di Fisica dell'Universit\`a  di Salerno and INFN, I-84084 Fisciano, Salerno, Italy \\
\PADOVA      . Dipartimento di Fisica dell'Universit\`a  di Padova, I-35131 Padova, Italy \\
\ANNECY      . LAPP, Universit\'e de Savoie, CNRS/IN2P3, F-74941 Annecy-le-Vieux, France \\
\BOLOGNA     . Dipartimento di Fisica dell'Universit\`a  di Bologna, I-40127 Bologna, Italy \\
\BOLOGNAINFN . INFN Sezione di Bologna, I-40127 Bologna, Italy \\
\FRASCATI    . INFN - Laboratori Nazionali di Frascati dell'INFN, I-00044 Frascati (Roma), Italy \\
\DUBNA       . JINR-Joint Institute for Nuclear Research, RUS-141980 Dubna, Russia \\
\LNGS        . INFN - Laboratori Nazionali del Gran Sasso, I-67010 Assergi (L'Aquila), Italy \\
\NAPOLI      . Dipartimento di Scienze Fisiche dell'Universit\`a Federico II di Napoli, I-80125 Napoli, Italy \\
\BARIINFN    . INFN Sezione di Bari, I-70126 Bari, Italy \\
\LAQUILA     . Dipartimento di Fisica dell'Universit\`a dell'Aquila and INFN, I-67100 L'Aquila, Italy \\
\STRASBOURG  . IPHC, Universit\'e de Strasbourg, CNRS/IN2P3, F-67037 Strasbourg, France \\
\HAMBURG     . Hamburg University, D-22761 Hamburg, Germany \\
\MOSCOWITEP  . ITEP-Institute for Theoretical and Experimental Physics, RUS-117259 Moscow, Russia \\
\MUNSTER     . University of M\"unster, D-48149 M\"unster, Germany \\
\NAGOYA      . Nagoya University, J-464-8602 Nagoya, Japan \\
\HAIFA       . Department of Physics, Technion, IL-32000 Haifa, Israel \\
\URBINO      . Universit\`a degli Studi di Urbino "Carlo Bo", I-61029 Urbino, Italy \\
\FUNABASHI   . Toho University, J-274-8510 Funabashi, Japan \\
\ZAGREB      . IRB-Rudjer Boskovic Institute, HR-10002 Zagreb, Croatia \\
\GAZWADONG   . Gyeongsang National University, ROK-900 Gazwa-dong, Jinju 660-300, Korea \\
\KARIYA      . Aichi University of Education, J-448-8542 Kariya (Aichi-Ken), Japan \\
\ROMA        . Dipartimento di Fisica dell'Universit\`a  di Roma ``La Sapienza" and INFN, I-00185 Roma, Italy \\
\BARI        . Dipartimento di Fisica dell'Universit\`a  di Bari, I-70126 Bari, Italy \\
\UTSUNOMIYA  . Utsunomiya University, J-321-8505 Tochigi-Ken, Utsunomiya, Japan \\
\ROSTOCK     . Fachbereich Physik der Universit\"at Rostock, D-18051 Rostock, Germany \\
\BRUSSELS    . IIHE, Universit\'e Libre de Bruxelles, B-1050 Brussels, Belgium \\
\NOWLNGS     . Now at INFN - Laboratori Nazionali del Gran Sasso, I-67010 Assergi (L'Aquila), Italy \\
\NOWINAFIASF . Now at INAF/IASF, Sezione di Milano, I-20133 Milano, Italy \\
\NOWROMA     . Now at Dipartimento di Fisica dell'Universit\`a  di Roma ``La Sapienza" and INFN, I-00185 Roma, Italy \\
\NOWPUSAN    . Now at Pusan National University, Geumjeong-Gu Busan 609-735, Korea \\
\NOWFRASCATI . Now at INFN - Laboratori Nazionali di Frascati dell'INFN, I-00044 Frascati (Roma), Italy \\
\NOWASAN     . Now at Asan Medical Center, 388-1 Pungnap-2 Dong, Songpa-Gu, Seoul 138-736, Korea \\
\NOWSUBATECH . Now at SUBATECH, CNRS/IN2P3, F-44307 Nantes, France \\
\NOWBERN     . Now at Albert Einstein Center for Fundamental Physics, Laboratory for High Energy Physics (LHEP),
               University of Bern, CH-3012 Bern, Switzerland \\
\CORR          Corresponding Authors: \\
               Email addresses: bertolin@pd.infn.it (A. Bertolin), 
               cecile.jollet@cern.ch (C. Jollet), \\
               anselmo.meregaglia@cern.ch (A. Meregaglia) \\
}

\newcommand{\OperaAuthorList}{
N.~Agafonova$^{\MOSCOWINR}$, % 1
A.~Aleksandrov$^{\NAPOLIINFN}$, % 2
O.~Altinok$^{\ANKARA}$, % 3
A.~Anokhina$^{\MOSCOWSINP}$, % 4
S.~Aoki$^{\KOBE}$, % 5
A.~Ariga$^{\BERN}$, % 6
T.~Ariga$^{\BERN}$,
D.~Autiero$^{\LYON}$, % 7
A.~Badertscher$^{\ZURICH}$, % 8
A.~Bagulya$^{\MOSCOWLPI}$
% ...
\\
}

\begin{document}

\pagestyle{empty} 

\title{
{\bf Study of neutrino interactions with \\
\vspace{0.2 cm}
the electronic detectors of the OPERA experiment} \\
\vspace{1cm}
OPERA COLLABORATION
}

\maketitle
\thispagestyle{empty} 

\author{\noindent \\ \OperaAuthorList }

\begin{flushleft}
\footnotesize{\OperaInstitutes }
\end{flushleft}

% \linenumbers

\vspace{0.5cm}

\begin{abstract}
\noindent
The OPERA experiment is based on a hybrid technology combining electronic
detectors and nuclear emulsions.
OPERA collected muon-neutrino interactions
during the 2008 and 2009 physics runs
of the CNGS neutrino beam, produced at CERN with an energy range of 
about 5-35 GeV.
A total of $5.3 \times 10^{19}$ protons on target equivalent luminosity has been
analysed with the OPERA electronic detectors: 
scintillator strips target trackers and magnetic muon spectrometers 
equipped with resistive plate gas chambers and drift tubes, allowing a detailed 
reconstruction of muon-neutrino interactions. 
Charged Current (CC) and Neutral Current (NC) interactions are identified,
using the measurements in the electronic detectors,
and the NC/CC ratio is computed.
The momentum distribution and the charge of the
muon tracks produced in CC interactions are analysed.
Calorimetric measurements of the visible energy are performed for both the
CC and NC samples.
For CC events the Bjorken-$y$ distribution and the hadronic shower profile are computed.
The results are compared to a detailed Monte Carlo simulation of the electronic
detectors' response. 
\vspace{1cm}
%\noindent{\it Keywords:} neutrino experiments  \\
\end{abstract}

\newpage
\pagestyle{plain} 
\setcounter{page}{1}
\setcounter{footnote}{0}

%%%%%%%%%%%%%%%%%%%%%%%%%%%%%%%%%%%%%%%%%%%%%%%%%%%%%%%%%%%%%%%%%%%%%%%%%%%%%%%
\section{Introduction}
\label{sec:intro}

OPERA~\cite{OPERAProposal} is a hybrid experiment based on Electronic 
Detectors (ED) and nuclear emulsions. It is exposed to the long-baseline 
CNGS beam~\cite{CNGS} from CERN to the Gran Sasso underground laboratory
(LNGS) 730 km away from the neutrino source. The main purpose of the 
experiment is the observation of $\nu_{\mu}$ to $\nu_{\tau}$ oscillations 
in the direct appearance mode. The $\nu_{\tau}$ are identified through 
the measurement of the $\tau$ leptons produced in their Charged Current 
(CC) interactions. The neutrino runs started in 2008 and a first $\nu_{\tau}$ 
candidate has recently been observed~\cite{Agafonova:2010dc}.
The beam is mainly composed of $\nu_{\mu}$; interactions due to the
$\bar{\nu}_{\mu}$, ${\nu}_{e}$ and ${\bar{\nu}}_{e}$ contamination
amount to 2.1~\%, 0.80~\% and 0.07~\% of the $\nu_{\mu}$ CC event rate.
In the following sections the $\nu_{\mu}$ interactions collected in
the 2008 and 2009 runs, corresponding to $5.3 \times 10^{19}$ protons
on target (p.o.t.)
are analysed with fully operating ED,
which have taken data for more than 98~\% of the active beam time.

The ED are of many uses in the OPERA analysis flow besides their crucial 
role in the trigger, in the location of the interaction point in the target volume
and in the muon identification process.
Thus, in this paper we discuss the OPERA ED performances in event 
selection, muon identification, momentum and charge reconstruction and
calorimetry measurements.

The main features of the OPERA ED are presented first, followed by a 
review of the MC simulation
and of the event reconstruction procedure.
CC and Neutral Current (NC) interactions measurements are then discussed.
The NC to CC ratio, the muon momentum spectrum, the reconstructed energy 
and the hadronic shower profile are investigated and a detailed 
comparison with a MC simulation is presented.
This analysis is also a benchmark to establish the quality of the MC 
simulation related to the ED.

%%%%%%%%%%%%%%%%%%%%%%%%%%%%%%%%%%%%%%%%%%%%%%%%%%%%%%%%%%%%%%%%%%%%%%%%%%%%%%%
\section{OPERA electronic detectors}

\begin{figure}[htbp]
\unitlength1cm  \begin{picture}(15.5,9.)
\includegraphics{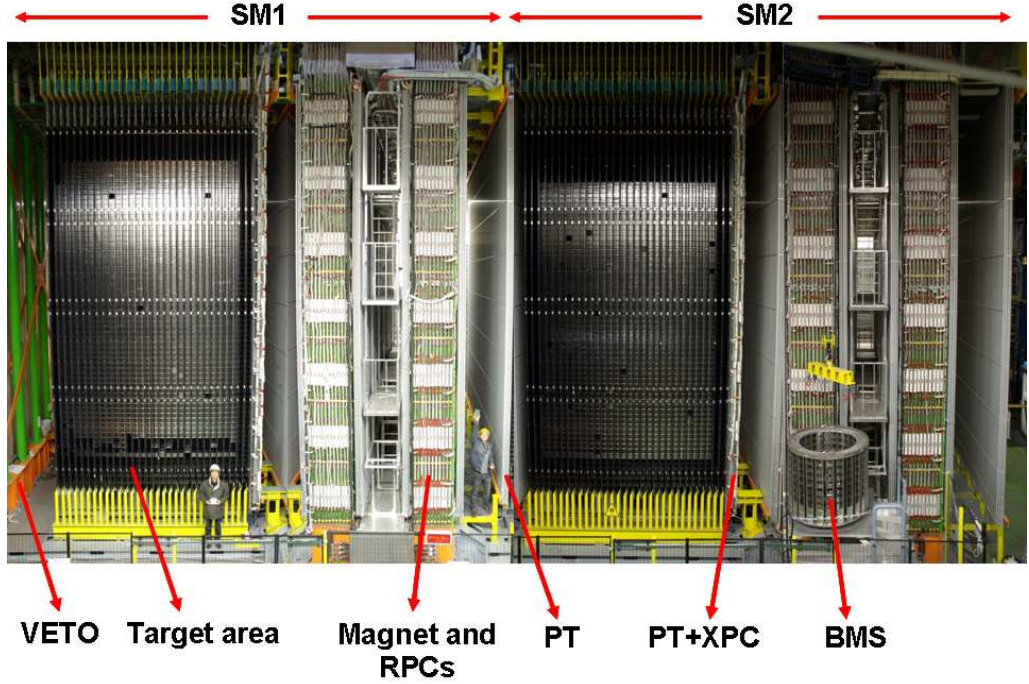}
\end{picture}
\vspace{-0.5cm}
\caption{View of the OPERA detector; the neutrino beam enters from the 
left. The upper horizontal lines indicate the two identical 
super-modules (SM1 and SM2). The target area is made of walls filled with 
lead/emulsion bricks interleaved with 31 planes of plastic scintillators 
(TT) per SM.
The VETO detector and a magnet with its inserted RPC planes are indicated 
by arrows, as well as some PT and XPC planes. 
The Brick Manipulator System (BMS) is also visible. 
See \protect{\cite{Detector}} for more details.}
\label{fig:opera_detector}
\end{figure}

As shown in figure~\ref{fig:opera_detector}, the OPERA detector~\cite{Detector} 
is composed of two identical super-modules (SM).
Each of them has a target section composed by target walls filled with
lead/emulsion bricks alternated with walls of scintillator strips that constitute 
the Target Tracker (TT).
Each target wall
contains about 2920 bricks and only 53 walls out of 62 are filled.
A brick is a mechanical unit which contains 57 emulsion films interleaved
with 56 1 mm thick lead plates. The transverse size of the brick is 12.8
$\times$ 10.2 cm$^2$. 
Each emulsion film has two 44 $\mu$m 
thick emulsion layers deposited on a 205 $\mu$m thick plastic base.
Each TT wall is composed of a pair of orthogonal
scintillator strip arrays with an effective granularity of 2.6 $\times$ 2.6
cm$^2$ and has a surface of 6.7 $\times$ 6.7 m$^2$ transverse to the
beam direction. Strips are read out via Wave Length Shifting (WLS) fibres 
connected to multi-anode photomultiplier tubes. 
The total masses of the lead/emulsion bricks and scintillator strips are 
about 1.25 and 0.3 kton, respectively.

A muon spectrometer at the end of each SM is used to identify muons and to 
measure their momentum and the sign of their charge.
Each spectrometer consists of a dipolar magnet with two arms made of 
12 iron plates;
the measured magnetic field strength is 1.52 T. 
The two arms are interleaved with 6 vertical drift-tube planes, the Precision
Trackers (PT), for the 
precise measurement of the bending of the muon tracks. 
Planes of Resistive Plate Chambers (RPC) are inserted between the iron 
slabs of the magnets, 11 planes in each arm.
Each RPC plane, 8.7 $\times$ 7.9 m$^2$ transverse to the beam 
direction, is equipped with two orthogonal sets of copper readout strips.
These planes provide a coarse tracking, a range measurement of the stopping 
particles and a calorimetric 
analysis of the hadrons escaping the target along the incoming neutrino
direction.
Two planes of resistive plate chambers (XPC), with the read out strips tilted by $\pm$
42.6$^{\circ}$ with respect to the horizontal, are also placed after each
target section to solve left/right 
ambiguities in the track pattern recognition.
Together with the RPC, the XPC are used to provide an external trigger to 
the PT.
A 10 $\times$ 9.12 m$^2$ anti-coincidence glass RPC detector, the VETO, is 
placed in front 
of the first SM to exclude (or tag) interactions occurring 
in the material and in the rock upstream of the target.
Although the ED are not conceived to perform calorimetric measurements, 
they can be used for this purpose with a coarse resolution.

\begin{figure}[htbp]
\unitlength1cm  \begin{picture}(15.5,9.)
\includegraphics{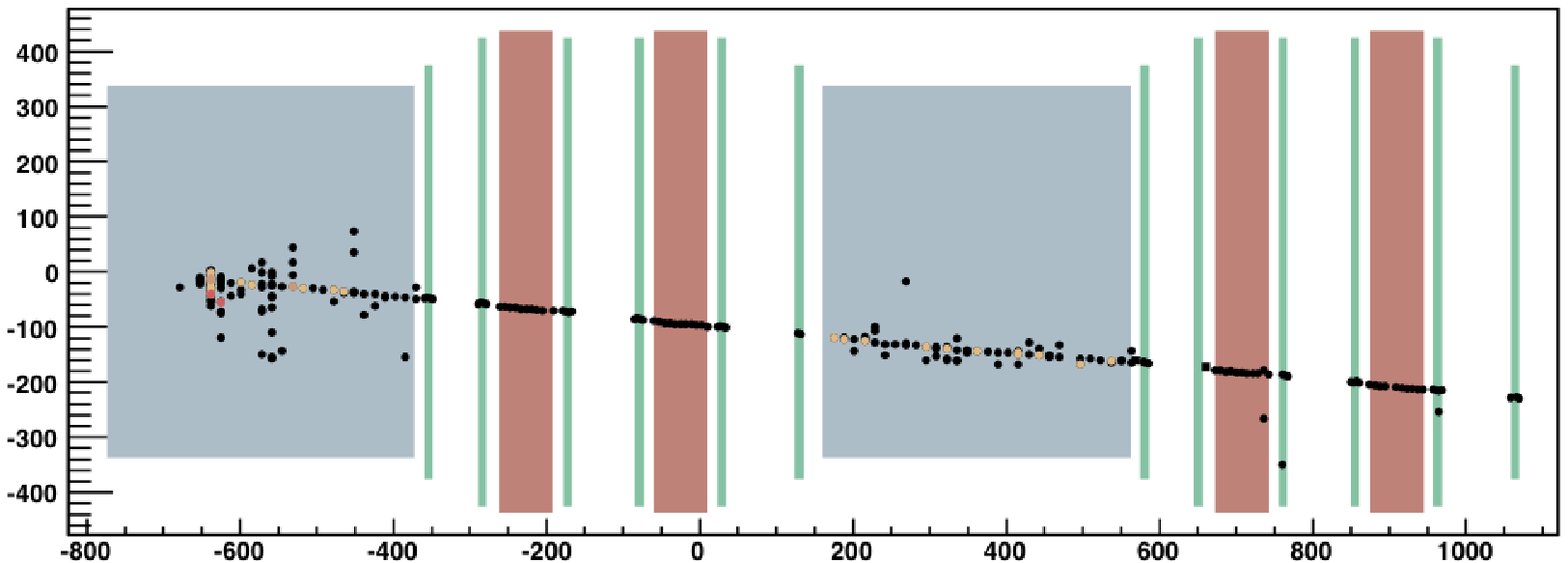}
\includegraphics{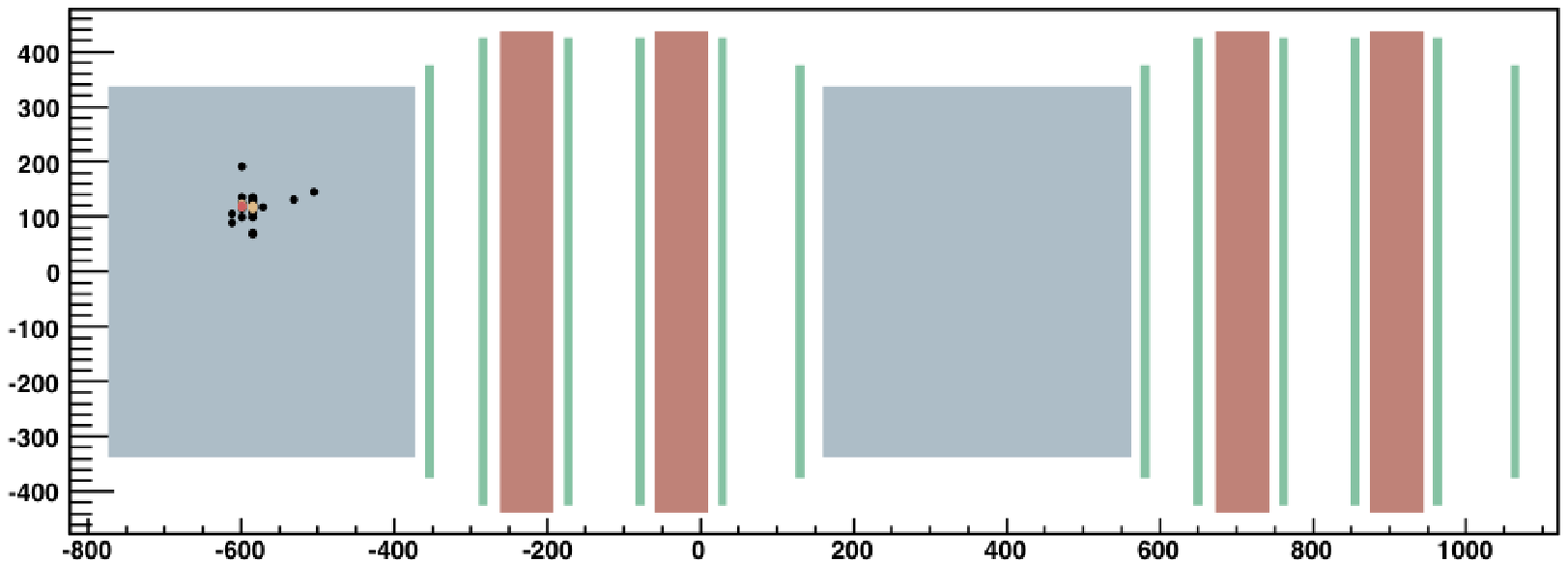}
\put(9.0,0.0){$z$ coordinate (cm)}
\put(9.0,4.4){$z$ coordinate (cm)}
\put(2.4,1. ){\begin{rotate}{90} $x$ coordinate (cm) \end{rotate}}
\put(2.4,5.5){\begin{rotate}{90} $x$ coordinate (cm) \end{rotate}}
\end{picture}
\caption{Examples of charged current (top) and neutral current (bottom) 
events as seen in one projection view of the OPERA electronic detectors.
In this view, the 2 SM can be recognised: for each one, the target is followed 
by the muon spectrometer.}
\label{fig:ccnc}
\end{figure}

An example of a CC event as seen by the OPERA ED is shown in the top
part of figure~\ref{fig:ccnc}, where the long tail of hits easily identifies
a very high momentum muon track. The bottom part of figure~\ref{fig:ccnc} shows
a NC event.
The connection between the ED and the nuclear emulsion data is described
in~\cite{Agafonova:2010dc,opera:njp8,Anokhina}.

%%%%%%%%%%%%%%%%%%%%%%%%%%%%%%%%%%%%%%%%%%%%%%%%%%%%%%%%%%%%%%%%%%%%%%%%%%%%%%%
\section{MC Simulations}

\subsection{Events Generation}
\label{sec:MC}

CC interactions can occur in the Quasi-Elastic (QE), Resonant (RES) or
Deep Inelastic Scattering (DIS) regimes.
In the QE and RES processes the hadronic system observable in the detector
is faint as most of the incoming neutrino energy is transferred to the final
state lepton. Conversely in the DIS process a prominent hadronic system
is observed in the detector.
In order to get a prediction for the number of expected neutrino 
interactions in OPERA, the differential neutrino cross 
sections, $d\sigma/dE$, for the CC-DIS, CC-QE and CC-RES processes on an
isoscalar target, known from other experiments~\cite{PDG}, are convoluted 
with the CNGS neutrino flux.
The mean energy of the incoming beam is 17.7~GeV~\cite{SL-Note-2000-063}
if the long tail extending above 100~GeV is not included.
Only 0.6 \% of the total flux has an energy 
exceeding 100~GeV, this corresponds to less than 4~\% of the $\nu_{\mu}$ CC
event rate on an isoscalar target~\cite{CNGS}.
Once the detector target mass and the number of recorded p.o.t. are 
defined, the absolute 
prediction for the expected number of interactions is computed together 
with the relative fractions of each process.
The CC-DIS, CC-QE and CC-RES fractions are corrected for the 
non-isoscalarity of the materials used in the OPERA 
detector.
The CC-DIS process is found to be dominant with a fraction exceeding 
90~\%.
Using the procedure outlined in reference~\cite{PDG}, the NC to CC 
ratio on an isoscalar target is predicted to be 0.31, while it is 0.29 once 
the non-isoscalarity corrections are taken into account.
Hence the NC contribution is also fixed. Only NC interactions 
in the deep inelastic scattering regime are considered 
as final state particles have to be observed in the ED.
The final states for the different processes, CC-DIS, CC-QE, CC-RES and
NC-DIS, are generated using the NEG MC program~\cite{Autiero:2005ve}, developed 
in the framework of the NOMAD experiment~\cite{Altegoer:1997gv}.
The NEG MC is supplied with the CNGS neutrino spectrum up to 400 GeV.
The generated events can then be mixed according to the appropriate fractions. 

\subsection{Environment simulation of the OPERA detector}

For the event simulation in the target, neutrino interaction primary 
vertexes are generated 
in the lead/emulsion volumes as well as in the volumes of the
TT scintillator strips.
The lead represents 93~\% of the target mass, the rest 
being emulsion films.
Neutrino interactions do not occur only in the target but also in
the other OPERA detector structures, like the spectrometers,
in any material 
present in the experimental hall, including the BOREXINO~\cite{Alimonti:2000xc} 
detector and its related facilities, and in the surrounding rock.  
The ratio between the number of recorded interactions occurring inside 
and outside the target is about 1 to 6.
As a consequence, primary vertexes have been generated in all of the above 
mentioned volumes since final state particles from any of
these volumes may easily reach the OPERA target.
Due to the asymmetric beam energy profile, with a tail at very high 
energy, a large enough volume of rock has to be considered.
Upstream of the detector a cylinder of rock, 35~m of radius and 
300~m of length, has been used in the simulation.
The rock volume surrounding the detector has the shape of a cylinder
with a radius of 35~m and an inner empty volume corresponding to
the LNGS Hall C shape hosting the OPERA detector.
MC studies show that 99 \% of the external events with hits reaching
the OPERA detector are contained in a volume which is 35 \% smaller 
than the simulated one. 
Once primary vertexes are generated, the produced outgoing particles are 
propagated through the different simulated volumes and their interaction 
with matter, either with passive elements, like the rock, or with a 
sensitive detector volume, is performed using the GEANT3~\cite{geant3} 
Virtual MC simulation package, version 1.10.

%%%%%%%%%%%%%%%%%%%%%%%%%%%%%%%%%%%%%%%%%%%%%%%%%%%%%%%%%%%%%%%%%%%%%%%%%%%%%%%
\section{Event reconstruction}

In order to provide a comparison with the data, the reconstruction of 
the simulated neutrino interactions is performed,
using the same algorithms as for real data, and the efficiencies 
of the different analysis steps, such as the selection of
neutrino interactions with the primary vertex contained in the target,
NC vs CC event tagging or muon identification are evaluated. 

\subsection{Selection of neutrino interaction events inside the target}
\label{sec:opcarac}

The OPERA DAQ system~\cite{Detector,Marteau:2009ct} records with
high efficiency all the interactions leaving a significant 
activity in the OPERA detector.
To achieve this the ED data are actually acquired in triggerless mode 
since the read out of the front-end electronics is asynchronous with the 
data, time stamped with a 10 ns clock. A minimum bias filter is applied at the
level of subdetectors in order to reduce the detector noise. The event 
building is then performed by collecting all the hits in a sliding time 
window of 3000 ns and requiring: hits in the $x$ and $y$ projections of at least 
two TT planes or a TT plane with the sum of the photomultiplier signals exceeding 
1500 ADC counts, and the presence of at least 10 hits.
If a muon track is present in the final state, the trigger efficiency of the DAQ 
system, estimated with MC methods, exceeds 99~\%.
Even in the worst configuration where a $\nu_\mu$ to $\nu_\tau$ oscillation occurs 
followed by a QE $\nu_\tau$ interaction and a $\tau$ to $e$ decay,
the trigger efficiency, averaged over the $\nu_\mu$ energy spectrum, exceeds 
95~\%.

Cosmic ray induced events are also recorded~\cite{OperaCos} but they can be 
easily rejected as they are not on time with the CNGS beam.
Therefore an almost pure sample of 31576 interactions on time with
the beam, with the primary vertex contained inside or outside the OPERA 
target, was obtained for the 2008-2009 neutrino runs.
As events occurring in the target represent a small fraction of the total number of 
recorded events an automatic algorithm, OpCarac~\cite{OpCarac}, 
identifies such events, called hereafter ``contained" events.
The contained events are more precisely those located in the target volume actually filled 
with bricks, this volume is fully instrumented by the TT, the walls of which being 
larger than the target walls.
Events not fulfilling this requirement are called 
hereafter ``external" events.
The OpCarac algorithm efficiency, estimated as the ratio between the number 
of MC events generated in the target volume
and selected as contained to the total number of MC events generated,
is high, as shown in 
table~\ref{tab:car}, in particular for CC events.

The external events are mostly due to CC interactions occurring in the rock
surrounding the detector. The final state muon is crossing the full
detector or entering from the sides. These events are easily
identified and rejected by OpCarac.
The presence of the VETO system is particularly helpful when the muon is 
entering the detector from the front.
The number of recorded external events with this topology was compared to MC 
expectations. Data and MC are in agreement within the
10~\% error on the expected number of $\nu_{\mu}$ CC events due
to the uncertainty on the total beam flux and on the $\nu_{\mu}$ CC cross
section.
Further MC studies show that CC events occurring in the rock
surrounding the Hall C volume, in which the final state muon escapes detection,
can also generate secondary particles produced at large angle with respect
to the incoming neutrino direction and hence reaching the target volume.
In this case the observed activity is mainly concentrated at the edges
of the TT. These events appear as low activity NC events.
Dedicated MC studies show
that the spatial distribution of the low activity NC events measured in
data and MC agree within the quoted 10~\% uncertainty.
Due to the low activity in the ED these events
are difficult to distinguish from genuine NC events occurring in the target.
The contribution of NC interactions outside the target volume
to the external event sample was checked through MC to be 20~\% of
the overall external sample.
In order to keep the efficiency for NC events occurring in the target high, a 
contamination of external events is unavoidable.
While the CC sample is basically free of external events, the 
contamination of the NC sample is at the 10~\% level and 3~\% for the whole
NC+CC sample. \\
\begin{table}[htbp]
\begin{center}
\begin{tabular}{|c|c|}
\hline
Type & Contained fraction \\
\hline
CC & 97.6 $\pm$ 1.4 \% \\
\hline
NC & 83.0  $\pm$ 1.6 \%\\
\hline
\end{tabular}
\end{center}
\caption{Efficiencies for the selection of contained events.}
\label{tab:car}
\end{table}%

\subsection{Muon Identification}
\label{sec:muid}

The muon identification performed by the ED is of primary importance in 
the OPERA analysis flow because:
\begin{itemize}
\item The $\tau$ muonic decay is a ``golden channel" to tag the $\nu_{\mu}$ 
to $\nu_{\tau}$ oscillation since it is the only channel where the momentum 
and charge of the decay daughter can be measured.
\item The identification of a muon track originating from the primary vertex
is of crucial importance to discard all the $\nu_{\mu}$ CC inclusive 
interactions which are a source of background for the $\tau$ search.
\item The muon charge measurement allows to discriminate muons coming from 
$\tau$ decay, with negative charge, from those produced by the decays of 
charmed particles, with positive charge. This background is unfavourably
large as charm is produced in $~\sim 4\%$ of CC interactions and the charm
branching ratio into $\mu^+$ is $~\sim 18\%$~\cite{PDG}.
\end{itemize}
According to the requirements defined in the OPERA detector proposal, a CC
tagging efficiency or similarly a muon identification efficiency greater than 95 \%
has to be attained.
Two algorithms have been developed~\cite{Notamuid}: the first one is based
on global event topology and is therefore independent from the track reconstruction 
efficiency. It can be applied to all the events, it is used for an evaluation of the 
NC/CC ratio, and it provides also a general veto for NC events.
The second algorithm relies on the muon track reconstruction and it can therefore 
be applied only to events where a track exists; it is mandatory for the connection 
of the muon track between ED and emulsions.

\begin{figure} [htbp]
\begin{center}
\mbox{\epsfig{file=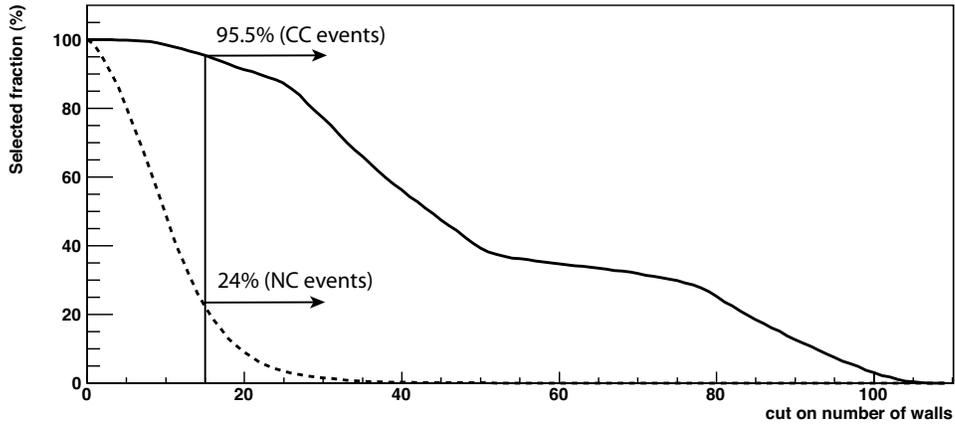,width=0.90\textwidth}} 
\caption{Integral fraction of selected events as a function of the cut on $N_{ED}$ 
for MC CC events (solid line) and MC NC events (dashed line). 
For this figure the contained event requirement is not applied.}
\label{fig:fraction}
\end{center}
\end{figure}

\begin{figure} [htbp]
\begin{center}
\mbox{\epsfig{file=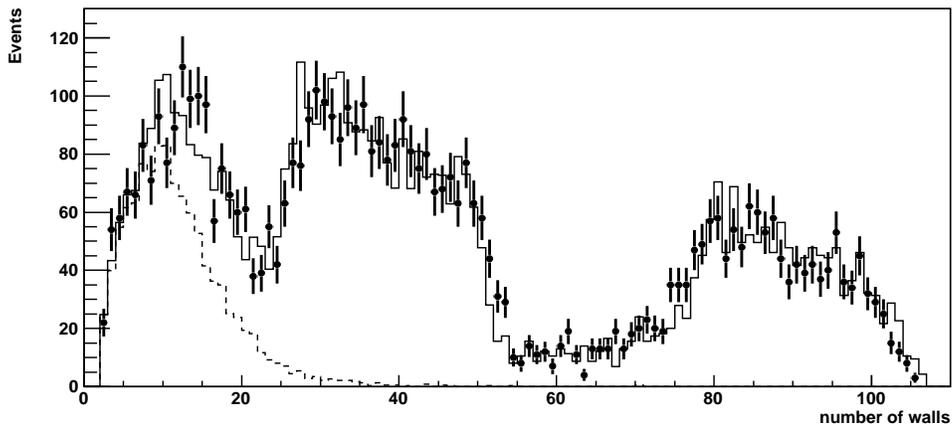,width=0.90\textwidth}} 
\caption{Number of hit walls for data (dots with error bars) and MC (solid line) 
contained events. The first bump is mainly due to NC events (dashed line) whereas 
the second and the third ones originate from CC events crossing one and two SM, 
respectively. The MC distribution has been normalised to data.}
\label{fig:walls}
\end{center}
\end{figure}

In the first algorithm the criterion to classify CC 
and NC events is based on the total number of ED planes containing hits ($N_{ED}$).
For the TT subdetector, $N_{ED}$ is obtained by counting the number of walls 
with signals in either of the two orthogonal planes; for the 
RPC subdetector the number of planes with signals in either of the two 
orthogonal sets of readout strips is considered.
The TT walls and the RPC planes are equally treated in this 
calculation.
The energy lost by a minimum ionising particle between two consecutive RPC planes 
is 57.1~MeV. Between two TT walls it is 71.4~MeV, 25 \% larger. The 
corresponding numbers of interaction lengths are respectively 0.298 and 0.328, 
10 \% larger.
The ratio between the $dE/dx$ and the interaction length of the two media is not so large 
to justify a different treatment in the algorithm applied.
Furthermore this difference is present both in data and MC.
In order to meet the requirement of the OPERA detector proposal of a CC
tagging efficiency greater than 95 \%, the lower cut on $N_{ED}$ must be set to 
14 planes, as can be seen in figure~\ref{fig:fraction}.
Correspondingly a large contamination of true NC events wrongly tagged as CC can not 
be avoided. MC studies showed that starting from a pure sample of NC events, 
about 24 \% are erroneously tagged as CC, 6\% relative to the full sample.
Events with $N_{ED} \leqslant$ 14 planes are instead tagged as NC. The tagging
efficiencies are summarised in table~\ref{tab:eff}.
The $N_{ED}$ distributions for data and MC events are shown in 
figure~\ref{fig:walls}, the agreement is reasonable.
\begin{table}[t]
\begin{center}
\begin{tabular}{|c|c|}
\hline
Type & Correctly identified fraction \\
\hline
CC & 95.5 $\pm$ 1.4 \% \\
\hline
NC & 76.0 $\pm$ 1.2 \% \\
\hline
\end{tabular}
\end{center}
\caption{MC efficiencies for CC and NC selection using the cut on $N_{ED}$. If  
$N_{ED} > 14$ the event is classified as CC, otherwise it is classified as NC. }
\label{tab:eff}
\end{table}

\begin{figure} [t]
\begin{center}
\mbox{\epsfig{file=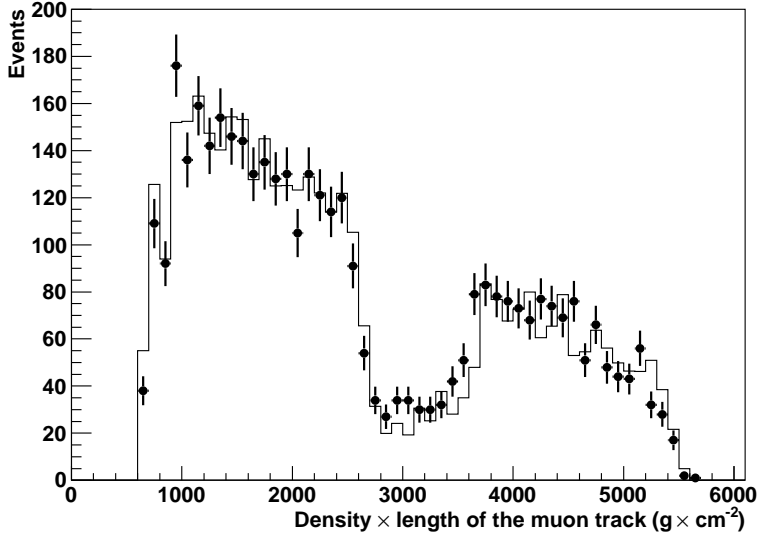,width=0.7\textwidth}} 
\caption{Length $\times$ density comparison for data (dots with error bars) and MC 
(solid line) for events classified as CC (i.e. length $\times$ density $> 660$ 
g~$\times$~cm$^{-2}$). The MC distribution has been normalised to data.}
\label{fig:muid}
\end{center}
\end{figure}

The second algorithm is based on a precise reconstruction and 
identification of the muon track.
The track reconstruction relies on a sequence of algorithms. Primarily the ED hits
are connected to form tracks in both longitudinal projections by a pattern recognition 
algorithm. 
Then the 3-dimensional tracks are reconstructed by associating tracks in the two 
projections.
A Kalman filter~\cite{Kalman} is also used to calculate the momentum and to 
reject hits wrongly assigned to tracks.
The length times density is then computed and used to identify a generic track as a 
muon track. The length is defined by adding straight distances between consecutive 
TT and/or RPC hits (TT walls are spaced by about 13~cm and RPC planes by about 
7~cm) along the whole track length. The actual detector structure along the track path 
fixes the value to be used for density.
The muon identification criterion is based on a cut on the length times density 
of the longest reconstructed 3-dimensional track in the event.
Requiring a muon identification at the level of 95~\% implies a cut at 660 
g~$\times$~cm$^{-2}$.
The length times density distributions for data and MC, above the selection cut, 
are shown in figure~\ref{fig:muid}, where the MC distribution has been normalised 
to the data. 
The MC simulation reproduces well the data trend.
For each track identified as a muon by the length times density 
criterion, the algorithm provides an estimate of the momentum.
If the track stops in the target or leaves the target but does not fully
cross at least one spectrometer arm, the energy is measured by range and the charge
measurement is not available. 
MC studies show that the NC contamination of the sample of events with at least
one muon track is 5.2~\%.
In the sample of events where a spectrometer is crossed, the NC contamination is as 
small as 0.8~\%. In addition, if at least one muon with negative charge is required 
the contamination drops to 0.4~\%. 

\subsection{NC to CC ratio}
\label{sec:nc2cc}

In this study, the algorithm based on the number of ED planes with hits  is used for 
identifying muons and hence CC events.
Applying first the contained event selection algorithm and then the muon 
selection criterium on $N_{ED}$ to the data, 4332 events are classified as CC 
($81.4 \pm 2.8$ \%) and 989 events as NC ($18.6 \pm 0.7$ \%). 
This gives a NC to CC ratio of $0.228 \pm 0.008$.

The MC estimation of the NC to CC ratio has to take into account the efficiencies 
of both the contained events selection, see table~\ref{tab:car}, and the correct 
CC and NC event tagging, see table~\ref{tab:eff}.
Moreover, as already seen in section~\ref{sec:opcarac}, detailed studies on 
the event selection show that in the data there is a non-negligible 
contamination of external events. 
Scaled to the number of true CC events occurring inside the target, this
contamination was estimated to be 2.96 \% for the NC and 0.78 \% for the CC
samples.
The number of reconstructed CC and NC events can be estimated via the following 
equations:
{\small \begin{eqnarray}
\label{eq:nccc}
CC_{rec} = \epsilon^C_{CC} \times \epsilon_{CC} \times n_{CC} +  
           \epsilon^C_{NC} \times (1-\epsilon_{NC}) \times R_{NC/CC} \times n_{CC} + 
           0.0078 \times n_{CC}\\
\nonumber NC_{rec} = \epsilon^C_{CC} \times (1-\epsilon_{CC}) \times n_{CC} +  
                     \epsilon^C_{NC} \times \epsilon_{NC} \times R_{NC/CC} \times n_{CC} + 
                     0.0296 \times n_{CC} 
\end{eqnarray}}
where:
\begin{itemize}
\item $\epsilon^C_{CC}$ is the efficiency of the contained events selection
algorithm for the CC MC sample.
\item $\epsilon^C_{NC}$ is the efficiency of the contained events selection
algorithm for the NC MC sample.
\item $\epsilon_{CC}$ is the efficiency of the CC selection algorithm for the 
CC MC sample.
\item $\epsilon_{NC}$ is the efficiency of the NC selection algorithm for the 
NC MC sample.
\item $n_{CC}$ is the true number of CC MC events.
\item $R_{NC/CC}$ is the true NC/CC ratio.
\end{itemize}
Using the value of $R_{NC/CC}$ computed in section~\ref{sec:MC}, 0.29,  and those of 
the efficiencies given in tables~\ref{tab:car}
and~\ref{tab:eff}, the MC expectation for the NC to CC ratio comes to 0.257. \\
The efficiencies for the contained events selection and for the NC and CC tagging
are extracted from the MC simulation, each with a statistical uncertainty in 
the range of 1 \% to 2 \%, as shown in tables~\ref{tab:car}
and~\ref{tab:eff}. These errors are then numerically propagated in 
equation~\ref{eq:nccc} to obtain a statistical uncertainty of 0.018 on the MC NC to 
CC ratio.

Systematic uncertainties due to CC and NC events tagging can be estimated 
by changing the cut on $N_{ED}$.
Varying the cut from 10 to 25 planes, the maximal discrepancy between
data and MC on the NC to CC ratio is $\pm$ 0.019. \\
Another source of systematics comes from the uncertainties in the contained events
selection algorithm.
As it can be seen in table~\ref{tab:car}, 17 \% of the NC MC events are not selected. 
Out of these, 4.2 \% are discarded since they would not fulfill the trigger condition 
(see section~\ref{sec:opcarac}).
The remaining 12.8 \% of the events are in a transition region, with little activity
recorded in the detector. Conservatively a 50 \% error is assumed for this particular 
topology of NC events.
The same computation for the CC events gives about a 1 \% error.
This propagates into an error of $\pm$ 0.015 on the final result. \\
In order to check for additional systematic uncertainties due to the contained 
events selection algorithm, data and MC calculations for the NC to CC ratio are 
repeated using events with the primary vertex in the first and the second SM 
separately. No difference is found in either the simulation or the real data where 
the values obtained for the ratio agree within 1 sigma.  Therefore, a possible 
contribution to the systematic error of the measurement is negligible. \\
The last source of uncertainties is on the number of
external events which affect mostly the NC sample.
This uncertainty is obtained by inspecting
the agreement between data and MC in variables that are particularly sensitive to
the external background component such as the visible energy and the three
dimensional position of the events. While genuine NC events
are uniformly distributed inside the target, external events tend to be
more concentrated towards the edges. 
This analysis showed that the expected number of background
events in data and MC are in reasonable agreement, within an uncertainty safely
estimated to be in the range of $- 15$ \% to $+ 24$ \%.
After numerical propagation in equation~\ref{eq:nccc}, this translates into an error 
of $\pm$ 0.006 in the final result.
Adding the different contributions in quadrature, the overall systematic uncertainty
on the NC to CC ratio for the MC is $\pm$ 0.025. \\
The results are shown in table~\ref{tab:NCCC} where statistical and systematic errors 
for MC have been added quadratically.
\begin{table}[t]
\begin{center}
\begin{tabular}{|c|c|}
\hline
NC/CC            & \\           
\hline
 Data & 0.228 $\pm$ 0.008 \\
\hline
 MC   & 0.257 $\pm$ 0.031\\
\hline
\end{tabular}
\end{center}
\caption{NC/CC ratio for data and MC.}
\label{tab:NCCC}
\end{table}

%%%%%%%%%%%%%%%%%%%%%%%%%%%%%%%%%%%%%%%%%%%%%%%%%%%%%%%%%%%%%%%%%%%%%%%%%%%%%%%%%%%
\section{OPERA ED performances on neutrino event reconstruction}

In the following paragraphs, data MC comparisons are presented on 
several reconstructed quantities characterising neutrino interaction 
events.

\subsection{Muon momentum reconstruction}
\label{sec:mumom}

\begin{figure}[t]
\begin{center}
\mbox{\epsfig{file=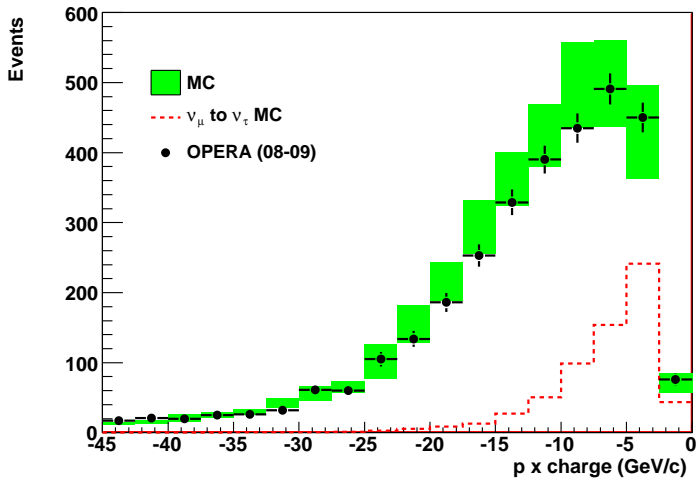,width=0.7\textwidth}} 
\caption{Muon momentum comparison (momentum $\times$ charge).
Data are shown by dots, errors are statistical only. 
The MC prediction, normalised to the number of p.o.t. 
corresponding to the 2008-2009 data sample, is shown by
the coloured area. The dominant source of the spread of the MC
prediction is due to the 10 \% uncertainty on the expected
number of $\nu_{\mu}$ CC events.
For illustration purposes only, the lower dashed curve represents 
the contribution obtained from the $\nu_{\mu}$ to $\nu_{\tau}$ MC 
events with subsequent decay into $\mu$ of the final state $\tau$ 
lepton. 
The normalisation of this contribution is arbitrary.}
\label{fig:mumom}
\end{center}
\end{figure}
A first step in establishing the quality of the muon momentum 
reconstruction can be done by comparing the momentum distribution
measured in data and MC for the contained events.
For this test it is desirable
to use a sample with a reduced NC contamination in order to disentangle 
the true muon reconstruction from possible effects due to background hadron 
tracks.
Therefore all events are required to have their muon momentum measured from 
the bending in the spectrometer.
In addition, a negative measured charge is required.
The measured muon momentum distribution is shown in figure~\ref{fig:mumom} 
and compared to MC expectations.
The MC has been normalised to the number of p.o.t. corresponding to the
2008-2009 data sample.
The error on the MC prediction is obtained after taking into account the 
uncertainty in the value of the magnetic field, which translates into a 3~\% 
shift of the MC spectrum, and the already quoted 10~\% uncertainty on the expected 
number of $\nu_{\mu}$ CC events,
the latter being the dominant source of uncertainty.
The contribution obtained from the $\nu_{\mu}$ to $\nu_{\tau}$ MC events with 
subsequent decay into $\mu$ of the final state $\tau$ lepton is also
shown in figure~\ref{fig:mumom}, with an arbitrary normalisation, to
show the interesting momentum region.
The spectrum of $\mu$ from $\tau$ decay is much softer than the
spectrum measured for $\nu_{\mu}$ CC interactions: the mean values obtained from the
MC simulation are -6.8 and -12.7 GeV/c, respectively.
In order to perform a shape comparison both data and MC distributions have
been normalised to 1. A $\chi^2$ value of 16.56 for 17 d.o.f. is obtained
without considering the magnetic field and the incoming neutrino flux 
uncertainties.
The overall normalisation was also checked: the number of events in data and 
MC agree within 6~\%, well within uncertainties.

\subsection{Muon charge reconstruction}

\begin{figure}[htbp]
\begin{center}
\mbox{\epsfig{file=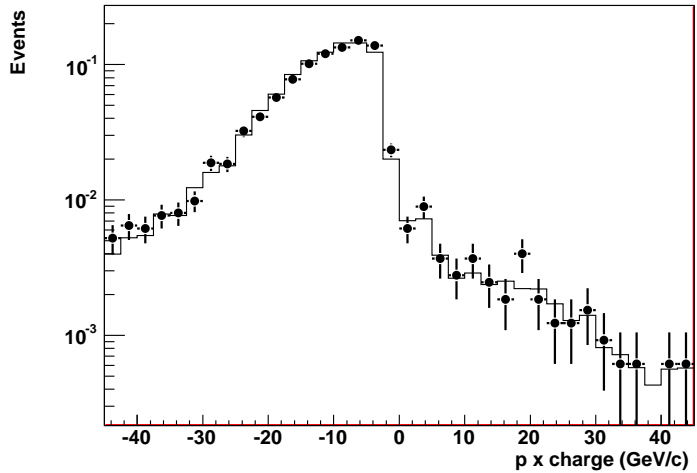,width=0.7\textwidth}} 
\caption{Muon charge comparison (momentum $\times$ charge): data
(black dots with error bars) and MC (solid line) are 
normalised to one.}
\label{fig:mucharge}
\end{center}
\end{figure}

As mentioned in section~\ref{sec:intro}, a $\bar{\nu}_{\mu}$ component
is present in the beam, leading via the CC process to positive muon tracks.
These can be used to test the muon charge reconstruction
algorithm by performing a measurement of the $\mu^+$ to $\mu^-$ events ratio.
The efficiency of the algorithm has been studied on CC 
MC events. It is defined as the fraction of simulated muon tracks reconstructed 
with the true charge.
As expected, the charge determination uncertainty increases with the muon momentum.
If an upper limit on the absolute value of the momentum is set at 45~GeV/c
the wrong determination of the muon charge is smaller than 2~\%.
The charge reconstruction
efficiency is also reduced at low momentum. In this case the
3-dimensional track identified as a muon may be
a charged hadron and hence the measured charge is not that of the muon.
This was not observed in MC events with the final state 
including a muon and negligible hadronic activity, as a confirmation
of this hypothesis.
Once again, if a lower limit on the absolute value of the momentum is set at
2.5~GeV/c, the wrong determination of the muon charge is smaller than 2~\%.
For muon momenta between 2.5~GeV/c and 45~GeV/c, the 
fraction of events with wrong charge determination is 1.2~\%.
The $\mu^+$ to $\mu^-$ events
ratio, within the selected momentum range, obtained from data can be 
directly compared with predictions based on the full MC sample: 
3.92 $\pm$ 0.37 (stat.) \% for
data, 3.63 $\pm$ 0.13 (stat.) \% for MC.
Figure~\ref{fig:mucharge} shows the momentum times charge distribution for 
data and MC, both normalised to one: the $\chi^2$ value is 23.34 for 35 d.o.f.

\subsection{Energy reconstruction}

This section studies the energy reconstructed using the 
TT subdetector.
A signal is measured at each end of the scintillator strips in terms 
of ADC counts (see~\cite{Adam:2007ex} for details), and then converted into a 
number of photo-electrons (p.e.) according to the gain of the photomultiplier 
(PMT) channel. 
The sum of the number of p.e. measured on both sides is converted into energy 
deposit (in MeV), according to the position of the hit along the strip and to a 
calibration curve that accounts for the attenuation of the signal along the
strip fibre. This 
calibration has been performed using radioactive sources before detector 
assembly and cosmic ray data. 
First the dependence of the number of p.e. of a minimum ionising particle (mip) 
on the crossing position along the strip has been validated 
(see section~\ref{sec:mip}), and used to compute the visible energy.
Then a calibration of the ED has been
performed in order to convert the visible energy into total energy
(see section~\ref{sec:RecE}).
The reconstruction algorithms are used to study, in data and MC, the 
distributions of the Bjorken-$y$ variable 
(section~\ref{sec:BY}).

\subsubsection{Visible energy}
\label{sec:mip}

\begin{figure} [htbp]
\begin{center}
\mbox{\epsfig{file=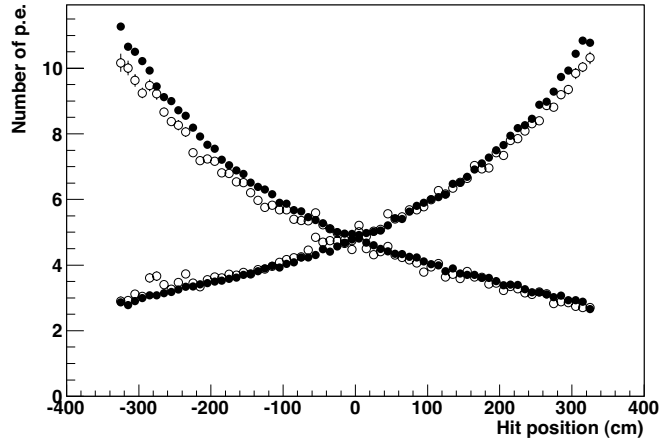,width=0.6\textwidth}} 
\caption{Number of detected p.e. on each extremity of the TT strips as a function of 
the hit position with respect to the left and right PMTs. The full circles are data, the empty ones are MC expectations.}
\label{fig:peTT}
\end{center}
\end{figure}

\begin{figure}[htbp]
\begin{minipage}{.49\textwidth}
\begin{center}
\mbox{\epsfig{file=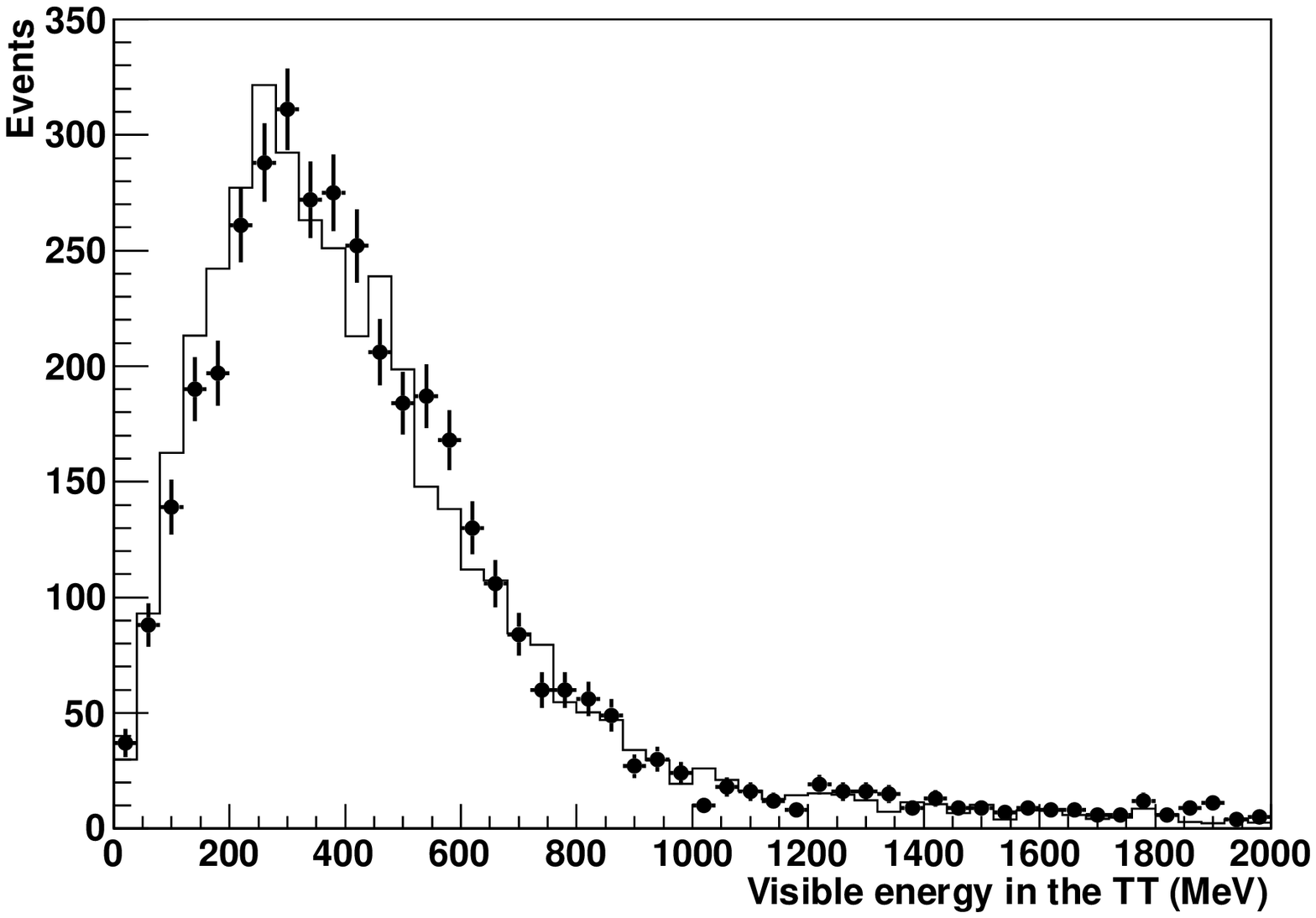,width=\textwidth}}
\end{center}
\end{minipage}
\hfill
\begin{minipage}{.49\textwidth}
\begin{center}
\mbox{\epsfig{file=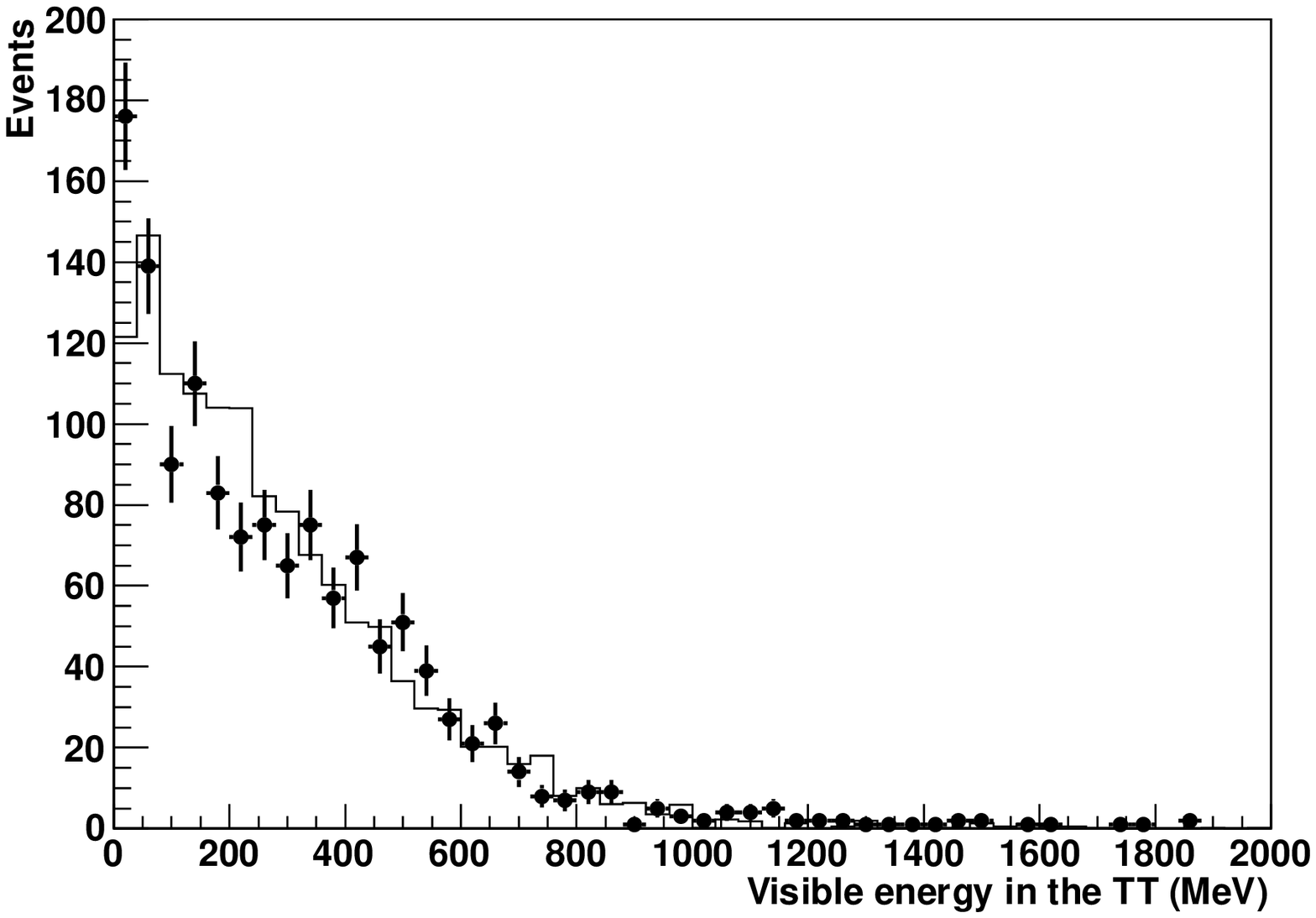,width=\textwidth}}
\end{center}
\end{minipage}
\caption{Energy deposit in the TT for events with at least one reconstructed
muon (left) and events with no muon (right). Dots with error bars correspond to 
data and solid lines to MC. MC distributions are normalised to data.}
\label{fig:ETT}
\end{figure}

Events with long tracks left by a mip have been selected and hits associated
only to those tracks have been considered. 
In figure~\ref{fig:peTT}, the number of p.e. recorded on each side of the fibre 
is plotted as a function of the distance to the hit. A double exponential decrease 
fits both the data and MC. The number of p.e. recorded at the centre of the fibre 
is typically 5.
The maximal discrepancy between data and MC is within 10~\%.
The reverse relation is used for converting a number of p.e. into 
visible energy once the hit position is reconstructed. This has been done for
events with at least one muon identified and events without muon identified separately, 
the comparison between data and MC is shown in figure~\ref{fig:ETT}.
There is on average a reasonable agreement between data and MC simulation, however some
discrepancies can be seen at very low energy for NC events. 
The energy deposit for NC events has been studied discarding soft NC-like events,
i.e. requiring at least one 3-dimensional reconstructed track, and the low 
energy disagreement disappeared. 

\subsubsection{Reconstructed energy}
\label{sec:RecE}

\begin{figure} [htbp]
\begin{center}
\mbox{\epsfig{file=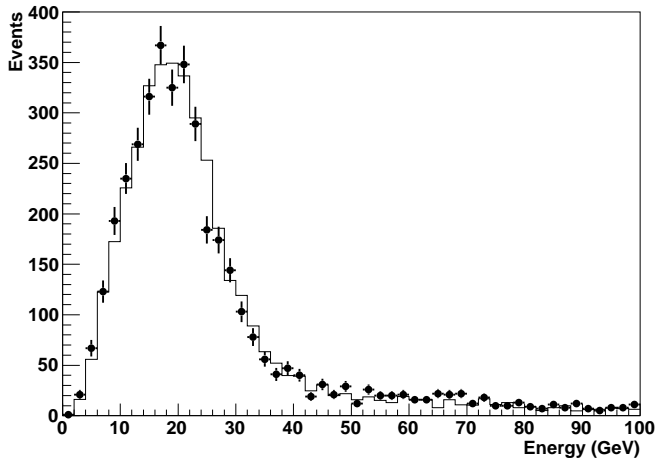,width=0.6\textwidth}} 
\caption{Total reconstructed energy for events with at least one identified muon
for data (dots with error bars) and MC (solid line). The MC distribution is normalised
to data.}
\label{fig:totE}
\end{center}
\end{figure}

In order to reconstruct the kinematical variables of an 
interaction, the knowledge of the total hadronic energy is required.
Based on a MC simulation, the relation between the true hadronic energy 
and the visible energy deposited in the TT and the RPC strips has been 
parametrised. The reverse parametrisation is used to estimate the 
hadronic energy from the ED data. 
Details on the energy resolution can be found in section~\ref{sec:appendix:tt}.
The results obtained by adding the hadronic and the muon energy are shown 
in figure~\ref{fig:totE} for events with at least one identified muon: 
data and MC are in reasonable agreement.

\subsubsection{Bjorken-$y$}
\label{sec:BY}

\begin{figure}[htbp]
\begin{minipage}{.49\textwidth}
\begin{center}
\mbox{\epsfig{file=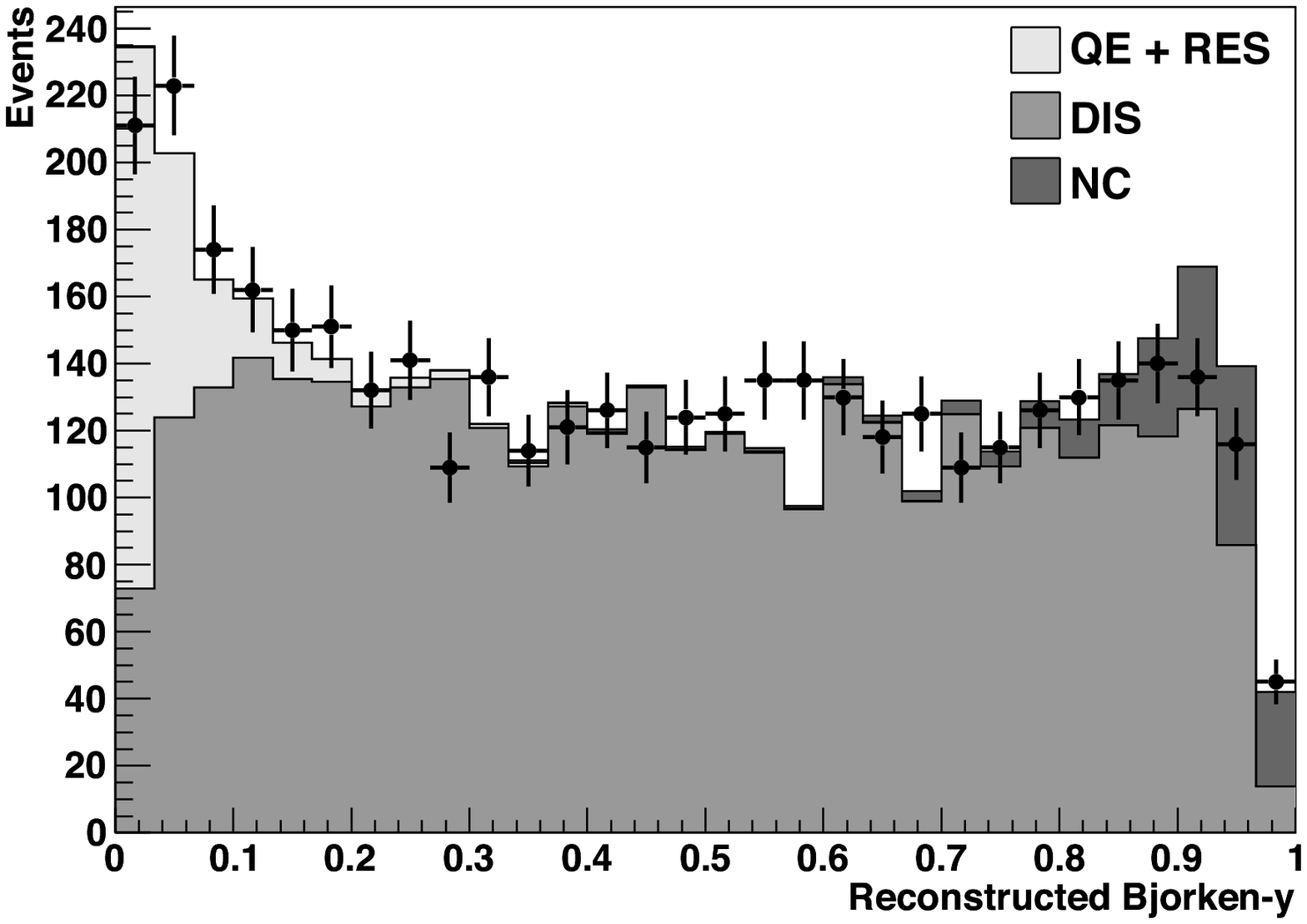,width=\textwidth}}
\end{center}
\end{minipage}
\hfill
\begin{minipage}{.49\textwidth}
\begin{center}
\mbox{\epsfig{file=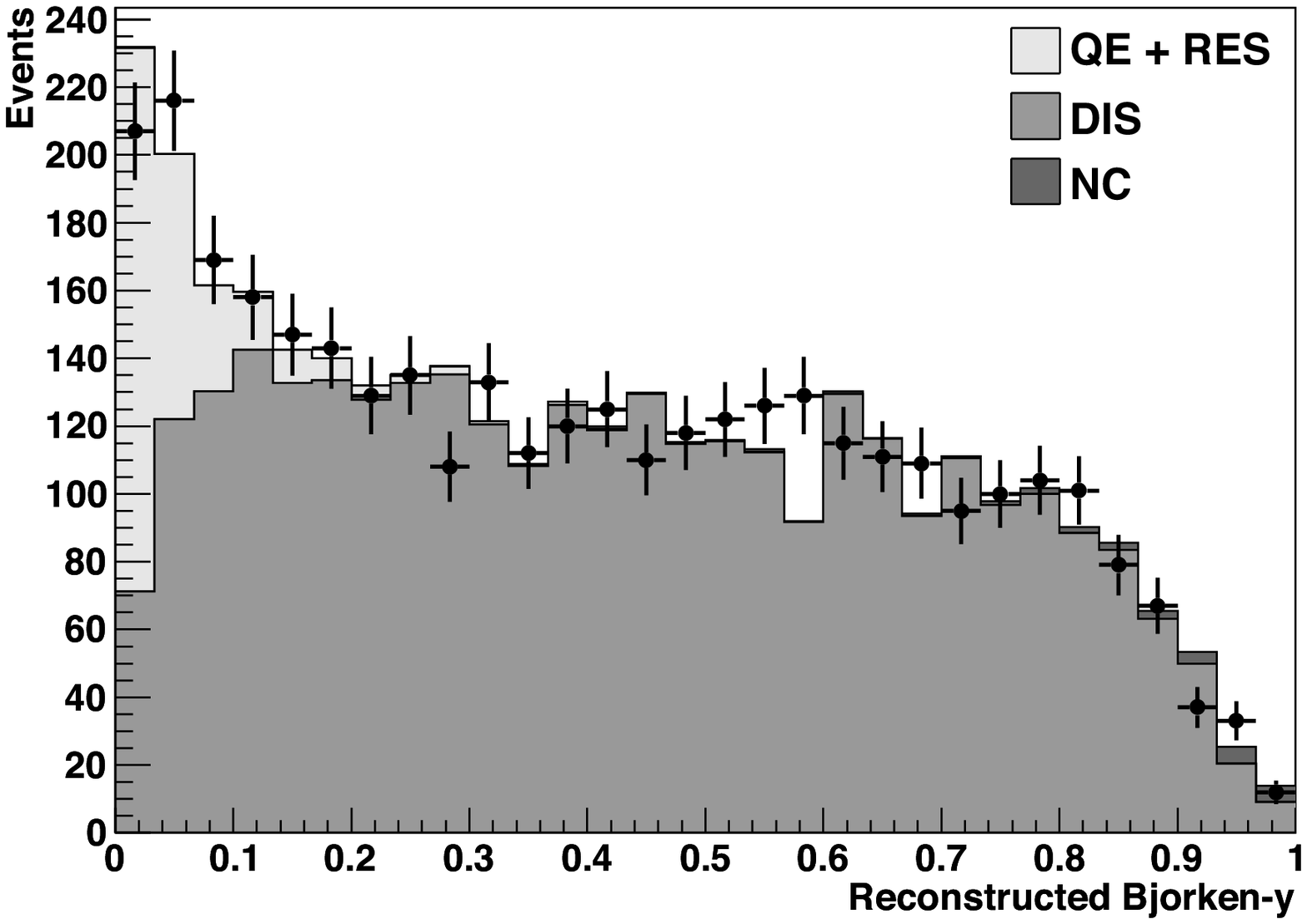,width=\textwidth}}
\end{center}
\end{minipage}
\caption{Bjorken-$y$ variable reconstructed in data (dots with error bars) and MC 
(shaded areas). The MC distributions are normalised to data. 
The different contributions of the MC are shown in different colours: QE + RES
contribution in light grey, DIS contribution in grey and the NC contamination in 
dark grey. 
On the left, all the events with at least one muon are shown whereas on the right 
the events for which the momentum is measured in the spectrometer are shown.}
\label{fig:BY}
\end{figure}

The Bjorken-$y$ variable represents the fraction of the hadronic energy with 
respect to the incoming neutrino energy.
For CC interactions:
\begin{equation}
\nu_{\mu}(k) N(P) \rightarrow \mu(k^{\prime}) X
\label{eq:process}
\end{equation}
where $k$, $P$ and $k^{\prime}$ are the quadrimomenta of the particles involved, 
the Bjorken-$y$ variable is defined as:
\begin{equation}
y = \frac{P \cdot (k - k^{\prime}) }{P \cdot k}.
\end{equation}
In the laboratory frame the Bjorken-$y$ variable can be computed as:
\begin{equation}
y = 1 - \frac{E_{\mu}}{E_{\nu_{\mu}}} = \frac{E_{had}}{E_{\mu}+E_{had}}
\label{eq:yBJ}
\end{equation}
where $E_{\nu_{\mu}}$ is the incoming neutrino energy, $E_{\mu}$ the energy of the 
outgoing muon and $E_{had}$ the hadronic energy.
Bjorken-$y$ connects the muon momentum measurement, performed in the spectrometer or by range, with the calorimetric measurements of all the 
hadrons.
The results for the events selected with at least
a muon track and for the events with the muon momentum measured by the spectrometer are shown in figure~\ref{fig:BY} in the left 
and right plots, respectively.
The agreement between data and MC simulation is reasonable: the $\chi ^2$ values are 55.4 and 48.7 respectively, for 29 d.o.f.
The sum of the QE and RES processes can be clearly seen as a peak at low 
$y$ values.
The NC contribution shows up at values of Bjorken-$y$ close to one. 
Figure~\ref{fig:BY} shows that the NC contribution becomes 
negligible when a track with its momentum measured by the spectrometer is required. 
This analysis results in an overall cross check of the performances of the ED.

\subsection{Hadronic shower profile}
\label{sec:had}

\begin{figure}[htbp]
\begin{minipage}{.49\textwidth}
\begin{center}
\mbox{\epsfig{file=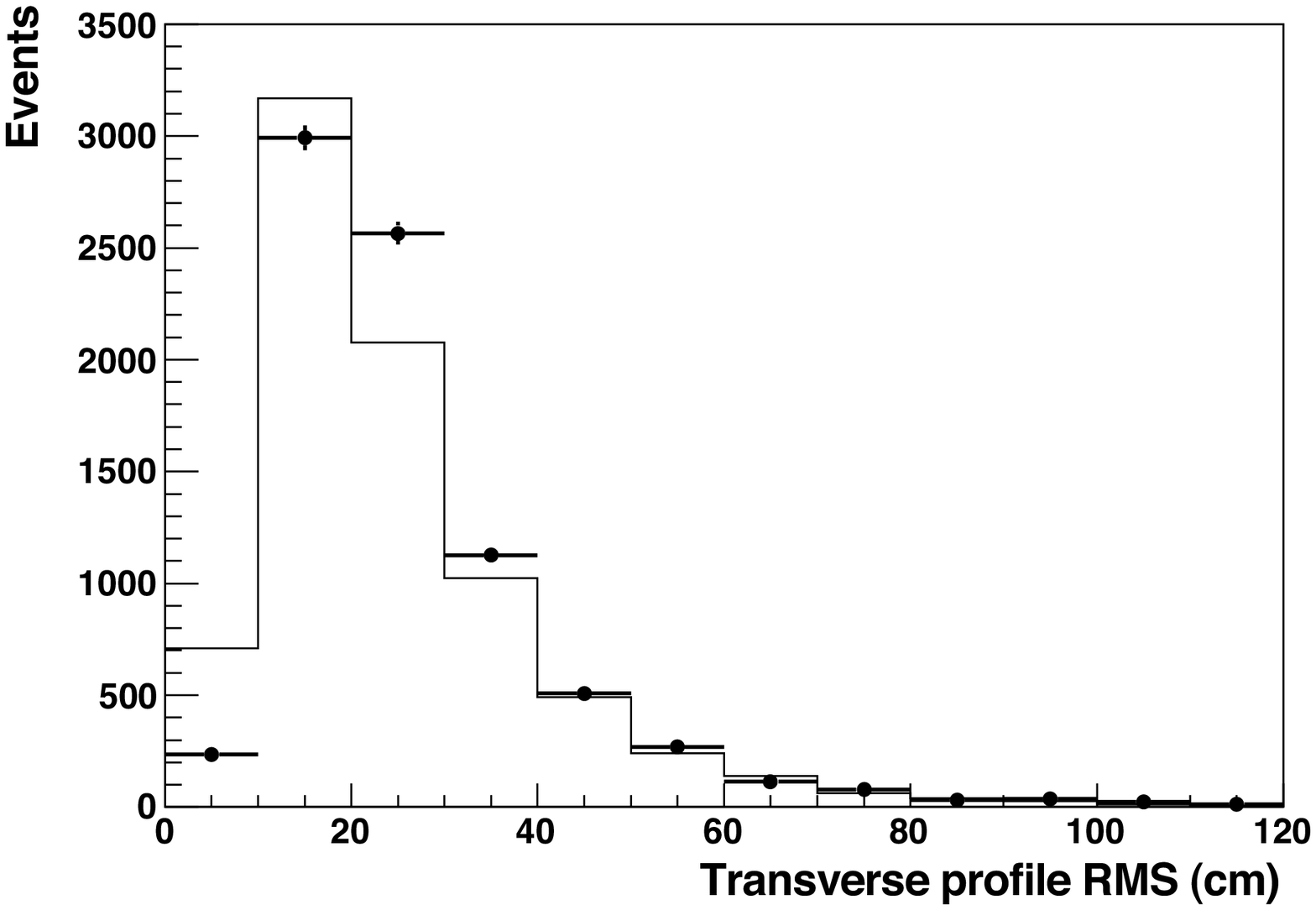,width=\textwidth}}
\end{center}
\end{minipage}
\hfill
\begin{minipage}{.49\textwidth}
\begin{center}
\mbox{\epsfig{file=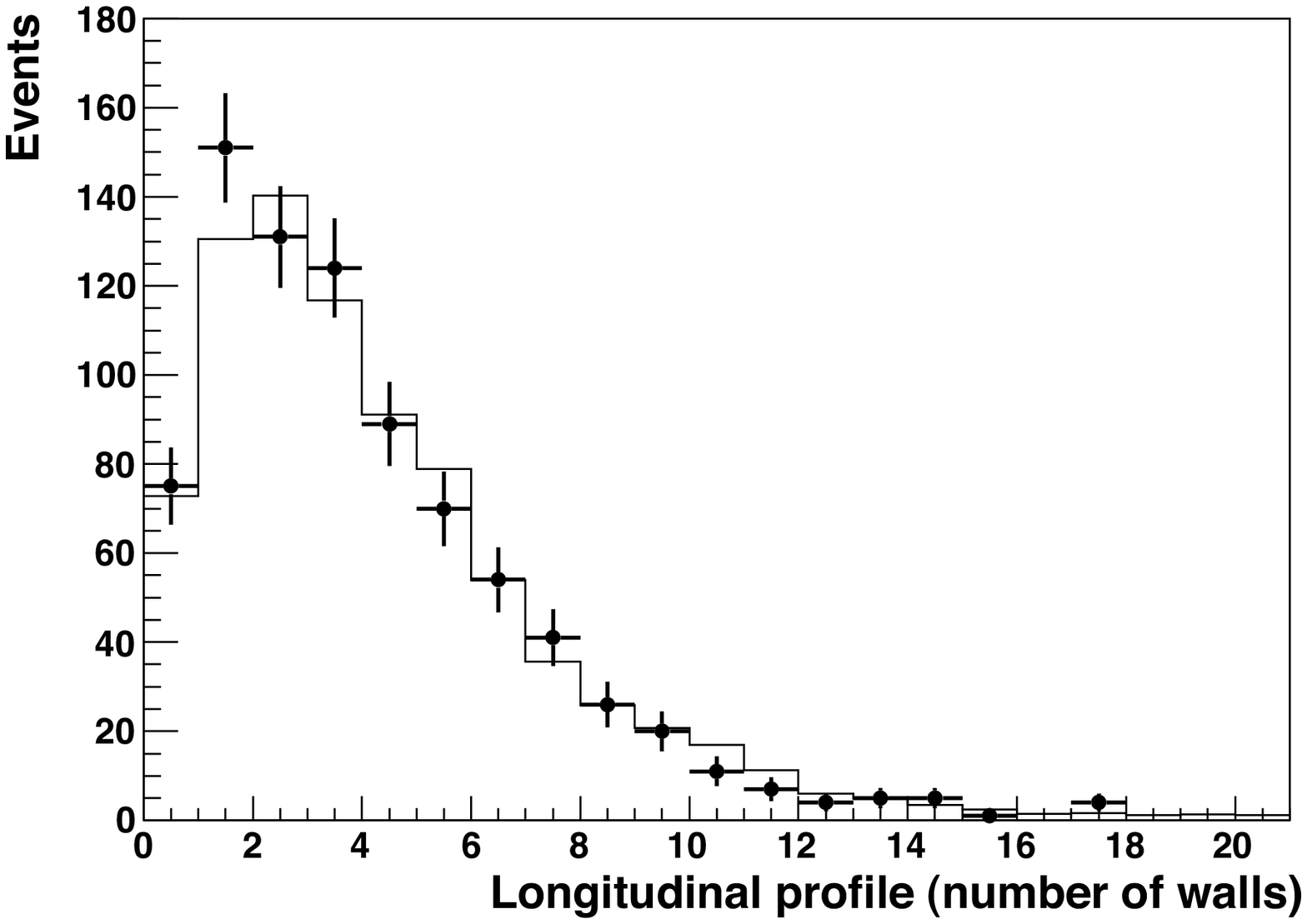,width=\textwidth}}
\end{center}
\end{minipage}
\caption{Transverse hadronic shower profile (left), in the X and Y projections, and longitudinal 
profile (right), in number of TT walls. 
Data are shown by dots with error bars and MC by the solid line. 
MC distributions are normalised to data.}
\label{fig:prof}
\end{figure}

A precise implementation in the MC simulation of the hadronic activity 
observed in data is very difficult: tools such as 
GHEISHA~\cite{gheisha} or FLUKA~\cite{fluka} describe imperfectly the 
measurements available.
Nevertheless, the hadronic activity is used, at least indirectly, in algorithms
such as the contained events selection or the brick finding.
Therefore, the hadronic shower profile in a sample of CC contained events has been analysed. 
The selected variables are the rms of the distribution of the shower profile in the 
X and Y projections (the transverse projections), where the TT hit positions 
are weighted by the number of collected p.e. 
The results are shown in figure~\ref{fig:prof} (left).
Similarly, the longitudinal profile of the shower is shown in figure~\ref{fig:prof} (right). 
In order to correctly calculate the longitudinal profile, the muon 
track has been removed, relying on an algorithm that finds the point where the muon 
exits from the shower and a clear track shows up.
Comparing the transverse profile the hadronic activity measured in data is broader
than in MC, 
whereas this effect is not visible in the longitudinal profile of the shower.
The simulation results shown here have been obtained with the GFluka option turned 
on in the GEANT3 simulation. 
The same comparison using the GHEISHA option yields a larger disagreement between data
and MC.

%%%%%%%%%%%%%%%%%%%%%%%%%%%%
\section{Conclusions}

The $\nu_{\mu}$ interactions data collected by the OPERA 
experiment in its two first running years, 2008 and 2009, have been
analysed using the full potentialities of the electronic 
detectors.
During this period, all electronic detectors were fully
operational for more than 98 \% of the active beam time.\\
Neutral and charged current interaction events have been analysed,
and a preliminary neutral to charged current event ratio has been 
measured and found to be consistent with MC expectations.
The analysis of the released energy profiles for both neutral and charged 
current events has been performed and a good agreement between data and MC
found. \\
Charged current events have been analysed in terms of muon momentum 
and charge reconstruction.
In particular, the muon charge ratio has been found to be consistent with the 
expected  $\bar{\nu}_{\mu}$ beam contamination. \\
In addition, charged current events have been used to measure the total 
reconstructed energy and the Bjorken-$y$ distribution.
Finally, the hadronic shower profile
has been analysed and compared to a MC simulation.
Although some characteristics of the energy profile are not reproduced
in detail the general description is satisfactory. \\
The results presented have shown the excellent performances of the OPERA 
electronic detectors and a good understanding of their simulation and response.

% Gaston Wilquet <Gaston.Wilquet@ulb.ac.be>: OK for Belgium
% Antonio Ereditato <antonio.ereditato@cern.ch>: for Switzerland
% "SNF, the Canton of Bern and the ETH Zurich for Switzerland"
% Mario Stip?evi? <Mario.Stipcevic@irb.hr>: no objections on our part
% Roganova Tatiana <rogatm@yandex.ru>: sentence added

\section{Acknowledgements}

We thank CERN for the commissioning of the CNGS facility and for its
successful operation, we thank INFN for the continuous support given to the
experiment during the construction, installation and commissioning phases
through its LNGS laboratory. We warmly acknowledge funding from our
national agencies: Fonds de la Recherche Scientifique - FNRS and Institut
Interuniversitaire des Sciences Nucl{\'e}aires for Belgium, MoSES for 
Croatia, CNRS and IN2P3 for France, BMBF for Germany, INFN for Italy,
JSPS (Japan Society for the Promotion of Science), MEXT (Ministry of
Education, Culture, Sports, Science and Technology ), QFPU (Global COE 
program of Nagoya University, ``Quest for Fundamental Principles in the
Universe" supported by JSPS and MEXT) and Promotion and Mutual Aid
Corporation for Private Schools of Japan, 
SNF, the Canton of Bern and the ETH Zurich for Switzerland,
the Russian Federal Property Fund, the grant 09-02-00300\_a, 08-02-91005-CERN,
Programs of the Presidium of the Russian Academy of Sciences ``Neutrino
Physics'' and ``Experimental and theoretical researches of fundamental
interactions connected with work on the accelerator of CERN'', Programs
of support of leading schools of thought (grant 3517.2010.2), Federal
Agency on a science and innovations state contract 02.740.11.5092 for
Russia, Department of priority directions of science and technologies,
the Korea Research Foundation Grant (KRF-2008-313-C00201) for Korea. 
We are also indebted to INFN for providing fellowships and grants
to non Italian researchers. We thank the IN2P3 Computing Centre (CC-IN2P3)
for providing computing resources for the analysis and hosting the central
database for the OPERA experiment. We are indebted to our technical
collaborators for the excellent quality of their work over many years of
design, prototyping and construction of the detector and of its facilities.
Finally, we thank our industrial partners.

\clearpage

%%%%%%%%%%%%%%%%%%%%%%%%%%%%%%%%%%%%%%%%%%%%%%%%%%%%%%%%%%%%%%%%%%%%%%%%%%%%%%%
\appendix
\setcounter{section}{1}
\section*{Appendix A: simulation of the Electronic Detectors}

In this appendix the simulation of the most relevant
subdetectors, TT, RPC and PT, will be reviewed in some detail.
The single strip, TT/RPC, or drift tube, PT, efficiencies 
implemented in the MC simulation are given in 
table~\ref{tab:effMC}.
The last subsection will also show the performances, in term 
of energy resolution, of the calorimetric measurements that 
can be achieved with OPERA. 
\begin{table}[htp]
\begin{center}
\begin{tabular}{|c|c|}
\hline
Detector & Efficiency \\
\hline
TT  & 99 \% (threshold at 1/3 p.e.)\\
\hline
RPC & 95 \%\\
\hline
PT  & 99 \% \\
\hline
\end{tabular}
\end{center}
\caption{Efficiency of the ED single strip or drift tube implemented 
in the MC simulation.}
\label{tab:effMC}
\end{table}%

\subsection{Simulation of the RPC}

When a particle is tracked through the volume occupied by the RPC
planes, one or more adjacent hit strips are created and the corresponding 
times are recorded.
Nine planes in each super-module are also used to generate a fast trigger
signal used as an external trigger by the drift tubes of the PT.
This trigger signal is also accurately computed in the simulation.
Due to the different width of the horizontal and vertical sets of 
readout strips, a slightly different efficiency is implemented in the MC 
simulation.
The efficiency in the MC is tuned from the observed multiplicity of
vertical and horizontal strips measured for neutrino induced muon tracks.
From these data and from cosmic ray data, the average strip efficiency
measured in situ exceeds 95 \%.
The stability of the performances is monitored using cosmic ray
data.

\subsection{Simulation of the PT}

If a charged particle passes through the gas filled volume of a PT drift 
tube, the hits in this volume are recorded as well as the corresponding 
time.
The hit nearest to the sense wire is taken to determine the drift time 
using a drift time to distance relation. In addition the drift time is smeared 
using a resolution function.
The time information of the RPC hits in planes contributing to the
trigger are used to generate a trigger time. The signal propagation delays
in all corresponding cables and in the RPC strips are taken into account
for a realistic simulation of the trigger time delay. 
The resulting trigger time is then subtracted from the time of the drift 
tube hit and this difference is used as an offset for the drift time. 
Thus trigger effects and the time of flight between the trigger planes and 
the drift tubes are properly accounted for. 
Also the signal propagation delay on the drift tube wires and the signal 
cables is taken into account.
In the simulation, the single tube efficiency is set to 99 \%. 
If no trigger is generated in a super-module, no drift tube data is saved for 
this super-module. 
The time to distance relation, the resolution function and the single tube 
efficiency have been determined using a test setup outside the LNGS, with 
the same operational parameters as used onsite.
Detector alignment is performed during dedicated cosmic ray runs without
magnetic field. In situ, performances are monitored using cosmic ray data.

\subsection{Simulation of the TT}
\label{sec:appendix:tt}

\begin{figure} [htbp]
\begin{center}
\mbox{\epsfig{file=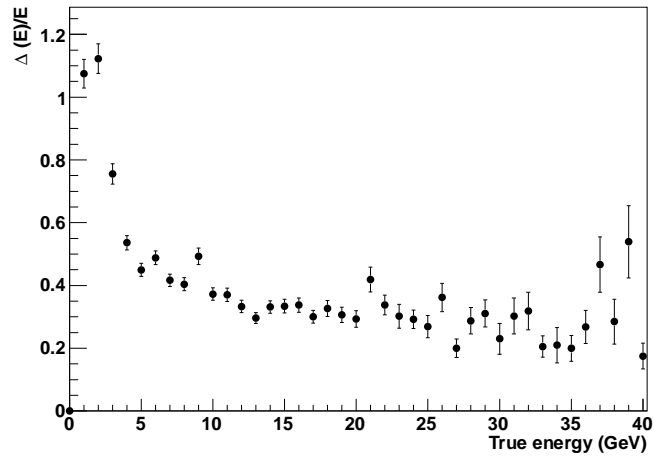,width=0.6\textwidth}} 
\caption{MC energy resolution computed using the CNGS neutrino energy spectrum.}
\label{fig:res}
\end{center}
\end{figure}

When a particle is tracked through the volume occupied by a TT
scintillator strip, the energy deposited and the time are recorded.
A corresponding light signal is generated, and the attenuation and
the delay in the propagation through the strip via the WLS fibre up to the corresponding
photomultiplier channel is computed. The signal is converted into a number of p.e. 
With the chosen threshold (1/3 of p.e.),
the mean detection efficiency for a minimum 
ionising particle crossing the strip in the middle is higher than 
99 \%. To make the detector description as realistic as possible the cross 
talk has also been included, i.e. the possibility that the 
signal deposited in one TT strip is recorded on a neighbouring 
photomultiplier channel.
Calibrations are periodically performed and efficiencies, obtained
from neutrino interaction data or cosmic ray data, are compared to
the MC simulation.
Using MC data, it is then possible to correlate the visible energy in
the TT with the incoming neutrino energy. In the presence of an energy
leakage from the TT into the spectrometer, the RPC data are also explicitly taken
into account by the algorithm.
The energy resolution reached is shown in figure~\ref{fig:res}.

\clearpage

%%%%%%%%%%%%%%%%%%%%%%%%%%%%%%%%%%%%%%%%%%%%%%%%%%%%%%%%%%%%%%%%%%%%%%%%%%%%%%%

\end{document}